\documentstyle[12pt]{article}
\newcommand{\be}{\begin{equation}}
\newcommand{\bea}{\begin{eqnarray}}
\newcommand{\eea}{\end{eqnarray}}
\newcommand{\ba}{\begin{array}}
\newcommand{\ea}{\end{array}}
\newcommand{\ee}{\end{equation}}

\expandafter\ifx\csname mathbbm\endcsname\relax

\else

\fi
\def\l{\label}

\begin{document}
\begin{titlepage}
\hfill
\vbox{
    \halign{#\hfil         \cr
           IPM/P-2003/022 \cr
           hep-th/0304247  \cr
           } 
      }  
\vspace*{20mm}
\begin{center}
{\Large {\bf On Effective Superpotentials and Compactification to 
Three Dimensions}\\ }

\vspace*{15mm}
\vspace*{1mm}
{Mohsen Alishahiha$^a$ \footnote{Alishah@theory.ipm.ac.ir}
and Amir E. Mosaffa$^{a,b}$}
\footnote{Mosaffa@theory.ipm.ac.ir} \\
\vspace*{1cm}

{\it$^a$ Institute for Studies in Theoretical Physics
and Mathematics (IPM)\\
P.O. Box 19395-5531, Tehran, Iran \\ \vspace{3mm}
$^b$ Department of Physics, Sharif University of Technology\\
P.O. Box 11365-9161, Tehran, Iran}\\

\vspace*{1cm}
\end{center}

\begin{abstract}
We study four dimensional ${\cal N}=2$ $SO/SP$ supersymmetric gauge theory
on $R^3\times S^1$ deformed by a tree level superpotential. We will show that
the exact superpotential can be obtained by making use of the Lax matrix of the 
corresponding integrable model which is the periodic Toda lattice. The
connection between vacua of $SO(2N)$ and $SO(2kN-2k+2)$ can also be seen in this
framework. Similar analysis can also be applied for $SO(2N+1)$ and $SP(2N)$.

\end{abstract}

\end{titlepage}

\section{Introduction}

It is believed that supersymmetric field theories have a rich
structure which make it possible to study some nonperturbative 
dynamics of the theory such as gaugino condensation or confinement
and chiral symmetry braking exactly (for a review, see e.g 
\cite{Intriligator:1995au}). This is mainly because 
one can find the exact form of 
the effective superpotential and thereby the structure 
of supersymmetric vacuum.
 
Recently an interesting step toward describing the vacuum structure,
as well as low energy coupling
for a wide class of ${\cal N}=1$ supersymmetric gauge theories was 
given by Dijkgraaf and Vafa \cite{Dijkgraaf:2002dh}. Motivated by earlier 
works \cite{Dijkgraaf:2002vw}-\cite{Bershadsky:1993cx} these authors
have conjectured that the exact superpotential and gauge coupling for
these ${\cal N}=1$ theories can be obtained using perturbative computations
in a related matrix model. Given an ${\cal N}=1$ SYM with a tree level 
superpotential, the potential of the corresponding matrix model is given in 
terms of the gauge theory tree level superpotential. This conjecture has been
verified using superspace perturbation formalism \cite{Dijkgraaf:2002xd}
or anomalies \cite{Cachazo:2002ry}.

More recently, in another attempt to study exact results in three dimensions,
the authors of \cite{Boels:2003fh} have considered ${\cal N}=2\; 
U(N)$ supersymmetric 
gauge theory deformed by a tree level superpotential on 
$R^3\times S^1$.\footnote{The compactification of the ${\cal N}=2$ SYM theory
to three dimensions was considered in \cite{Seiberg:1996nz}. For further
discussions see for example \cite{Katz:1996th}-\cite{Aharony:1997bx}.}
In fact, based on earlier works \cite{Dorey:1999sj}, 
they have observed that the vacuum structure of the theory can be
described using an integrable model that underlies the four dimensional theory
which in their case is the periodic Toda lattice based on
the root system of the 
affine Lie algebra $A_{N-1}^{(1)}$.\footnote{The relation
between ${\cal N}=2$ SYM theories and integrable system was discussed
in several papers including \cite{Gorsky:1995zq}-\cite{Mironov:2000se}.
For recent discussion in this direction and its relation with 
Dijkgraaf-Vafa conjecture see also \cite{Itoyama:2002rk}.} 
More precisely the authors of \cite{Boels:2003fh} have 
conjectured that if the classical superpotential is ${\rm Tr}W(\phi)$, 
then the quantum superpotential is given by ${\rm Tr}W(M)$, where $M$ 
is the Lax matrix of the corresponding integrable model. They have also 
shown that the structure of the supersymmetric vacuum can also be studied 
in this framework, in particular it was shown that the vacuum structure of 
$SU(N)$ gauge theory can be lifted to that of $SU(KN)$ gauge theory. 

It is the aim of this article to further investigate this proposal. 
In fact we will see that the proposal also works for ${\cal N}=2$ SYM theory
with $SO/SP$ gauge groups. In this case the integrable model is also
given in terms of the periodic Toda lattice,
though in the case of nonsimply laced group one needs to work with the dual
gauge group where the short roots and long roots are exchanged 
\cite{Martinec:1995by}.

The paper is organized as follows. In section 2 we shall review 
some field theory aspects of ${\cal N}=2$ $SO(2N)$ supersymmetric gauge 
theory on $R^3\times S^1$ and its reduction to three dimensions as well.
In section 3 we will demonstrate the way in which the effective 
superpotential can be obtained for ${\cal N}=2$ $SO(2N)$ supersymmetric 
gauge theory on $R^3\times S^1$ deformed by a tree level superpotential 
$W$. This has been done by making use of an integrable system.
In fact we will see that the Lax matrix of the corresponding integrable
model plays an essential role. In section 4 we shall show that the
lifting of supersymmetric vacua of $SO(2N)$ to $SO(2KN-2K+2)$ can also
be understood in the context of integrable model. In the section 5
the same scenario has been applied for $SO(2N+1)$ and $SP(2N)$ gauge groups. 
The last section is devoted to conclusion and comments.

\section{Field theory description}

\subsection{Three dimensional ${\cal N}=2\;\;SO(2N)$  SYM theory}
In this section we shall consider ${\cal N}=2$ $SO(2N)$ supersymmetric 
gauge theory in three dimensions. Having a gauge theory one should deal
with the vector multiplet of three dimensional SUSY theory.

The vector multiplet $V$, of three dimensional ${\cal N}=2$ supersymmetric
theory contains the gauge field, an adjoint scalar and two real
fermion gauginos which can be combined into a complex fermion. The
chiral/anti-chiral field strengths are defined by $W_{\alpha}=
-{1\over 4}{\bar D}^2e^{-V}D_{\alpha}e^V$ and ${\bar W}_{\alpha}=
-{1\over 4}D^2e^{-V}{\bar D}_{\alpha}e^V$, and the kinetic term
of the classical action is given by
\be
{1\over g^2_3}\int d^3xd^2\theta\; {\rm Tr}(W^{\alpha}W_{\alpha})+{\rm h.c.}\;.
\ee
  
The theory has a Coulomb branch where the real scalar $\phi$, which is
taken in the Cartan subalgebra of the gauge group, gets an
expectation value. In a generic point of the moduli space the gauge
group is broken to the Cartan subgroup where the gauge group is $U(1)^N$,
while at the boundaries of the Weyl chamber there is classically
enhanced gauge symmetry.

The $U(1)^N$ gauge fields, in the bulk of Coulomb branch, can be
dualized to scalars, $F_{\mu\nu}^a=\epsilon_{\mu\nu\rho}\partial^{\rho}
\sigma^a,\;a=1,\cdots N$. The $\sigma^a$ which parameterize the
Cartan torus of the gauge group, can be combined with the $\phi^a$
into chiral superfield $\Phi^a$ whose scalar components are $\phi^a+i\sigma^a$.
One can also think about this theory as compactification of four dimensional
${\cal N}=1$ gauge theory in which $\phi$ can be thought of as the component
of the gauge field in the reduced direction.

For three dimensional ${\cal N}=2$ gauge theory one expects to get a
superpotential. The nonperturbative contribution is due to three
dimensional instantons which are monopoles in the four dimensional theory.
These three dimensional instantons are associated with $\pi_2$ and since
$\pi_2(SO(2N))=0$ there can only be instantons in the Coulomb branch where
the gauge group is broken to $U(1)^N$ and one has $\pi_2(SO(2N)/U(1)^N)=Z^N$.
Therefore there are $N$ independent fundamental instantons associated
with the simple roots of the gauge group.
Their contribution to the nonperturbative superpotential is given by
\be
W=\sum_{i=1}^{N-1}e^{{\phi_i-\phi_{i+1}\over g_3^2}}
+ e^{{\phi_{N-1}+\phi_{N}\over g_3^2}}\;. 
\ee

\subsection{Four dimensional SYM theory on $R^3\times S^1$}

If we think about this three dimensional theory as a theory which
comes from reduction of a four dimensional ${\cal N}=1$
gauge theory on a circle of radius $R$ with $g_3^{-2}=Rg_{4}^{-2}$,
the superpotential develops an $R$ dependent term \cite{Seiberg:1996nz}
which in 
the case of $SO(2N)$ gauge theory is given by \cite{Katz:1996th}
\be
W=\sum_{i=1}^{N-1}e^{{\phi_i-\phi_{i+1}\over g_3^2}}
+ e^{{\phi_{N-1}+\phi_{N}\over g_3^2}}+ \gamma
e^{(-{\phi_{1}+\phi_{2}\over g_3^2})}\;. 
\ee
where $\gamma=e^{-1/Rg_3^2}$. From group theory point of view the last
term corresponds to the extra node one can add to the Dynkin diagram
to make an affine Dynkin diagram. The three dimensional theory
is obtained in the limit of $R\rightarrow 0$ while the four 
dimensional theory is recovered in large $R$ limit. 

The theory could also have a mass term for real scalar fields
$\phi_i$'s and therefore the whole superpotential reads
\be
W={1\over 2}m
\sum_{i=1}^{N}\phi_i^2+\sum_{i=1}^{N-1}e^{{\phi_i-\phi_{i+1}\over g_3^2}}
+ e^{{\phi_{N-1}+\phi_{N}\over g_3^2}}+ \gamma
e^{(-{\phi_{1}+\phi_{2}\over g_3^2})}\;,
\ee
which under a field redefinition can be recast to the following form
\be
W={1\over 2}m \sum_{i=1}^{N}\phi_i^2+m\sum_{i=0}^Ny_0\;,
\ee
where
\be
my_0=\gamma e^{(-{\phi_{1}+\phi_{2}\over g_3^2})},\;\;\;\;
my_N=e^{{\phi_{N-1}+\phi_{N}\over g_3^2}},\;\;\;\;
my_i=e^{{\phi_i-\phi_{i+1}\over g_3^2}}\;\;{\rm for}\;i=1,\cdots N-1\;.
\ee
We note, however, that there is a constraint on the variables $y_i$ which is
\be
\prod_{i=0}^Ny_i\prod_{j=2}^{N-2}y_j=\gamma m^{-(2N-2)}=\Lambda^{4N-4}\;,
\ee
with $\Lambda$ being the dynamical scale. Therefore one needs to impose this
constraint in the superpotential using a Lagrange multiplier $L$, as 
\be
W={1\over 2}m
\sum_{i=1}^{N}\phi_i^2+m\sum_{i=0}^Ny_0+L\log\left({\Lambda^{4N-4}\over
\prod_{i=0}^Ny_i\prod_{j=2}^{N-2}y_j}\right)\;.
\l{WMASS}
\ee

We could also consider the theory with a general classical superpotential
given by a holomorphic function $W(\Phi)$ on $R^3\times S^1$. 
Then the question would be how one can find the exact superpotential in this 
case. In the next section we will study this problem using the fact that 
this model is related to an integrable model.

\section{Quantum superpotential}

In this section we would like to study the exact superpotential
of ${\cal N}=1\;\; SO(2N)$ SYM theory on $R^3\times S^1$ which can be 
obtained from ${\cal N}=2$ Seiberg-Witten model deformed by
a classical superpotential $\int d^4xd^2\theta W(\Phi)$. To do this
we note that the Seiberg-Witten model is related to an integrable
system which is the periodic Toda lattice \cite{{Gorsky:1995zq},
{Martinec:1995by},{Nakatsu:1995bz}}.

The periodic Toda lattice associated to $SO(2N)$ gauge group is given
by the following Hamiltonian \cite{Bog}
\be
H={1\over 2}\sum_{i=1}^Np_N^2+\sum_{i=0}^{N}V_i\;,
\ee
where
\be\ba {ll}
V_{i}=\Lambda^2e^{q_i-q_{i+1}}\;\;\;\;\;&{\rm for}\;i=1,\cdots, N-1,\cr &\cr
V_{N}=\Lambda^2e^{q_{N-1}+q_{N}},\;\;\;\;\;&V_{0}=\Lambda^2e^{-(q_1+q_2)}\;,
\ea\ee
here $q_i$ are coordinates and $p_i$ are their corresponding momenta, and
$\Lambda^2$ is a parameter which will play the role of the dynamical
scale in the gauge theory side. This model is an integrable  model which 
means that there exists a Lax pair given by two matrices $M_{2N\times 2N}$ 
and $A_{2N\times 2N}$ that are functions of coordinates and momenta such that 
evolution of the theory can be described by the Lax equation
\be
{\partial M\over \partial t}=[M,A]\;.
\ee
The Lax matrix $M$, for $SO(2N)$ group is given by \cite{Adler}
\bea
M=\pmatrix{p_1 &V_1&&&&&&&-z&0 \cr 1&p_2&V_2&&&&&&&
z \cr
0&1&&&&&&&&\cr &&&&V_{N-1}&-V_N&&&& \cr
&&&1&p_n&0&V_{N}&&& \cr &&&-1&0&-p_N&-V_{N-1}&&& \cr
&&&&1&-1&-p_{N-1}&-V_{N-2}&& \cr &&&&&&-1&&& \cr 
-{V_{0}\over z}&&&&&&&&&-V_{1} \cr 0&{V_{0}\over z}&&&&&&&-1&-p_1}.\nonumber
\eea
We note also that there is a constraint on $V_i$'s, namely
$\prod_{i=0}^{N}V_i\prod_{i=2}^{N-2}V_i=\Lambda^{4N-4}$. To make a
connection with our discussion in the previous section it is useful to make
a change of variable in which $p_i=\phi_i$ and $V_i=y_i$. In particular
we note that
\be
{\rm Tr}(M^2)=4({1\over 2}\sum_{i=1}^N\phi_i^2+\sum_{j=0}^Ny_i)
\l{Todamass}
\ee 
which is the same as (\ref{WMASS}) up to a factor of two which is because of
$Z_2$ symmetry of the root system of $SO(2N)$. In fact this is a special 
case of the proposal made in \cite{Boels:2003fh}, namely the quantum
superpotential can be obtained from the classical one by replacing $\phi$ with 
the Lax matrix $M$
\be
\int dx^4d^2\theta\;W(\Phi)\rightarrow   \int dx^4d^2\theta\;W(M)\;.
\l{pro}
\ee
Of course to compare this with the gauge theory result one needs to identify
the integrable system parameters $(p_i,q_i)$ with the gauge theory fields
as what we have done in equation (\ref{Todamass}). Finally we note that the
spectral parameter $z$, does not appear in the quantum potential as long as we 
consider the classical superpotential with powers less than $2N-2$. For a
potential with a term of ${\rm Tr}(\phi^{2N-2})$ one finds a constant $z$ 
dependent term in the form of $(4N-4)(z+{\Lambda^{4N-4}\over z})$ which,
following \cite{Boels:2003fh}, we will simply drop!

\subsection{An example}
To see how the proposal works for the $SO(2N)$ gauge theory, let us
consider the 
theory with the gauge group $SO(8)$ with tree level superpotential 
\be 
W={g_2\over 2}\;{\rm Tr}(\phi^2)+{g_4\over 4}\;{\rm Tr}(\phi^4)\;.
\ee
The corresponding Lax matrix for gauge group $SO(8)$ is given by
\be
M=\pmatrix{\phi_1&y_1& 0&0&0&0&-z&0\cr 1&\phi_2 &y_2 &0 &0 &0 &0 &z \cr  
0&1 &\phi_3 &y_3 &-y_4 &0 &0 &0 \cr 0&0 &1 &\phi_4 &0 &y_4 &0 &0 \cr 
0&0 &-1 & 0&-\phi_4 &-y_3 &0 &0 \cr 0&0 &0 &1 &-1 &-\phi_3 &-y_2 &0 \cr 
-{y_0\over z}&0 &0 &0 &0 &-1 &-\phi_2 &-y_1 \cr 0&{y_0\over z} &0 &0 &0 &0
 &-1 &-\phi_1 }\;,
\ee
The Seiberg-Witten curve is obtained from the spectral curve given by 
$\det(x{\bf 1}-M)=0$ which reads\footnote{The Seiberg-Witten curve for 
$SO(2N)$ SYM theory was first obtained in \cite{Brandhuber:1995zp}.}
\be
x^2(z+{\Lambda^{12}\over z})-{1\over 4}(x^8-u_2x^6+\cdots+u_8)=0\;,
\ee 
where $y_2\prod_{i=0}^4y_i=\Lambda^{12}$.

Following \cite{Boels:2003fh} the effective superpotential reads
\be
W_{\rm eff}={g_2\over 2}\;{\rm Tr}(M^2)+{g_4\over 4}\;{\rm
Tr}(M^4)+2L\log\left(
{\Lambda^{12}\over y_2^2\prod_{i=0}^4y_i}\right) \;.
\l{effs}
\ee
Here we have also imposed the constraint on $y_i$'s using a Lagrange
multiplier $L$. 
The equations of motion for $\phi_i$'s are given by
\bea
W'(\phi_1)+g_4(y_1\phi_2-y_0\phi_2+2y_1\phi_1+2y_0\phi_1)&=&0\;,\cr 
W'(\phi_2)+g_4(y_2\phi_3+y_1\phi_1-y_0\phi_1+2y_1\phi_2
+2y_2\phi_2+2y_0\phi_2)&=&0\;,\cr 
W'(\phi_3)+g_4(y_3\phi_4-y_4\phi_4+y_2\phi_2+2y_3\phi_3
+2y_4\phi_3+2y_2\phi_3)&=&0\;,\cr 
W'(\phi_4)+g_4(y_3\phi_3-y_4\phi_3+2y_3\phi_4+2y_4\phi_4)&=&0\;,
\eea
where $W'(\phi_i)=g_2\phi_i+g_4\phi_i^3$, while the equations of motion for
$y_i$'s read
\bea
g_2+g_4(y_0+y_2+3y_1+\phi_2^2+\phi_1^2-\phi_1\phi_2)&=&Ly_0^{-1}\;,\cr
g_2+g_4(y_1+y_2+3y_0+\phi_2^2+\phi_1^2+\phi_1\phi_2)&=&L y_1^{-1}\;,\cr
g_2+g_4(y_2+y_1+y_3+y_4+y_0+\phi_3^2+\phi_2^2+\phi_2\phi_3)
&=&2L y_2^{-1}\;,\cr
g_2+g_4(y_3+y_2+3y_4+\phi_3^2+\phi_4^2+\phi_3\phi_4)&=&Ly_3^{-1}\;,\cr
g_2+g_4(y_4+y_2+3y_3+\phi_3^2+\phi_4^2-\phi_3\phi_4)&=&L y_4^{-1}\;.
\eea
Moreover the equation of motion of $L$ gives $y_2\prod_{i=1}^4y_i=\Lambda^{12}$.

Any solution of these equations would lead to a supersymmetric vacuum of the
theory. In particular one could consider solutions with $\phi_1=\phi_4=0$,
$\phi_2=\phi_3=\phi$
and $y_i=y$ for $i\neq 2$. In this case we are left with the following equations
for $\phi, y$ and $y_2$
\bea
W'(\phi)+g_4(4y+3y_2)\phi&=&0\;,\cr
g_2+g_4(y_2+4y+\phi^2)&=&{L\over y}\;,\cr
g_2+g_4(y_2+4y+3\phi^2)&=&{2L\over y_2}\;.
\eea
Let us first study a situation where $\phi=0$. For this case one finds
\be
\phi_i=0,\;\;\;\;y_2=2y,\;\;\;\;\;y_i=y\;\;{\rm for}\;i=0,1,3,4,
\;\;\;\;\;L=y(g_2+6g_4y)\;.
\ee
where $y={1\over 2^{1/3}}\Lambda^2$. This solution corresponds to the case 
where classically the gauge group remains unbroken and thus we get 
maximally confining vacuum.
 
Plugging this solution into the effective superpotential (\ref{effs}) one finds
\be
W_{\rm eff}=6(2^{2/3}g_2\Lambda^2+3\;2^{1/3}g_4\Lambda^4)\;.
\l{Weff}
\ee

Let us now assume that $\phi\neq 0$ then the solution is give by
\be
\phi^2=-({g_2\over 4g_4}+{5\over 2}y),\;\;\;\;\;y_2=-({g_2\over
4g_4}+{1\over 2}y),
\;\;\;\;\;y_i=2\;\;{\rm for}\;i=0,1,3,4\;.
\ee
where $y$ is obtained from the following equation
\be
{1\over 2}y^3+{g_2\over 4g_4}y^2+\Lambda^6=0\;,
\ee
which can be solved in power series of $\Lambda$. The result is
\be
y=-{g_2\over 2g_4}-8{g_4^2\over g_2^2}\Lambda^6+256{g_4^5\over
g_2^5}\Lambda^{12}-14336{g_4^8\over g_2^8}\Lambda^{18}+
983040\;{g_4^{11}\over g_2^{11}}\Lambda^{24}+
{\cal O}(\Lambda^{28})\;.
\ee
Therefore the effective superpotential (\ref{effs}) for this solution reads
\be
W_{\rm eff}={g_2^2\over g_4}\left(-1+16\;{g_4^3\over g_2^3}\;\Lambda^6-
128\;{g_4^6\over g_6^3}\;\Lambda^{12}+4096\;{g_4^9\over g_2^9}\;\Lambda^{18}
-196608\;{g_4^{12}\over g_2^{12}}\;\Lambda^{24}
\right)+{\cal O}(\Lambda^{28})\;. 
\l{rrr}
\ee
From the solution we found, one can see that this vacuum corresponds to
the case 
where the gauge group is classically broken as
 $SO(8)\rightarrow SO(4)\times U(2)$.

These results should be compared with the 4-dimensional field theory. To
do this
we note that the 4-dimensional field theory result could be obtained
using factorization
of the corresponding Seiberg-Witten curve. In general the factorization
problem is 
hard to solve, but for the maximally confining vacuum 
there is a general solution given by Chebyshev polynomials. For $SO(N)$ gauge 
group the solution is \cite{Edelstein:2001mw}
\be
\langle u_{2p}\rangle={{\tilde h}\over 2p}\pmatrix{2p\cr p}\Lambda^{2p}\;.
\l{SWf}
\ee
where ${\tilde h}$ is dual Coxeter number of the gauge group. Therefore in 
our model the effective superpotential reads
\be
W_{\rm eff}=g_2\langle u_{2}\rangle+g_4\langle u_{4}\rangle=
6(g_2\Lambda^2+{3\over 2}g_4\Lambda^4)
\ee
which is the same as (\ref{Weff}) if we rescale
$\Lambda \rightarrow 2^{1/3}\Lambda$. This rescaling can also be
understood from the 
spectral curve we have found using our notation for Lax matrix $M$. In fact
in comparison with the spectral curve in \cite{Martinec:1995by} we have
an extra 
factor of ${1\over 4}$ which can be absorbed by a redefinition of $z$ and 
$\Lambda$. Actually one needs to rescale $z\rightarrow z/4$ and
$\Lambda\rightarrow \Lambda/4^{(4/4N-4)}$ to make our notation as that in
\cite{Martinec:1995by}. 
Therefore in the case of $SO(8)$ one gets a rescaling factor of $2^{1/3}$.

In the case where the gauge symmetry is also broken as $SO(8)\rightarrow 
SO(4)\times U(2)$, the factorization of Seiberg-Witten can also be worked out.
In fact this model has been studied in \cite{Fuji:2002vv} where the effective
superpotential was given in a power series of $\Lambda$ as following 
\be
W_{\rm eff}={g_2^2\over g_4}\left(-1+4\;{g_4^3\over g_2^3}\;\Lambda^6-
8\;{g_4^6\over g_6^3}\;\Lambda^{12}+64\;{g_4^9\over g_2^9}\;\Lambda^{18}
-768\;{g_4^{12}\over g_2^{12}}\;\Lambda^{24}
\right)+{\cal O}(\Lambda^{28})\;,
\ee
which is the same as $(\ref{rrr})$ upon the rescaling of $\Lambda
\rightarrow 2^{1/3}\Lambda$, as expected.

Note that if we had considered a ${g_6\over 6}{\rm Tr}(\phi^6)$ term in the
classical superpotential one would have got the following quantum
superpotential
\be
W=6(2^{2/3}g_2\Lambda^2+3\;2^{1/3}g_4\Lambda^4+{40\over 3}g_6
\Lambda^6)+4g_6(z+
{\Lambda^{12}\over  z})\;,
\ee
which has a $z$ dependent term. We note also that the $\Lambda^6$ term is
indeed in agreement with the field theory result coming from (\ref{SWf}).
Therefore, as we mentioned before, it seems that the $z$ dependent term
has no contribution and should be dropped \cite{Boels:2003fh}, though we
do not have any strong physical reason for it.

\section{$SO(2KN-2K+2)$ gauge theory from $SO(2N)$ theory }

It has been shown \cite{Fuji:2002vv} that a solution for the massless monopole 
constraints of the $SO(2KN-2K+2)$ can be obtained from that of $SO(N)$
via Chebyshev
polynomials. Since the compactified theory is closely related to an integrable 
system one might wonder if such a structure could also be seen in this
framework.
In this section we shall examine this relation using their corresponding
integrable systems. In fact we will show that this information is indeed
encoded 
in the structure of the Lax matrix of both theories.
Actually our discussion in this section is parallel to
that in \cite{Boels:2003fh} where the lifting from $SU(N)$ to $SU(KN)$
was studied.

Consider the following $2KN\times2KN$ matrix made out of the Lax matrix 
of the $SO(2N)$ periodic Toda lattice
\bea
&&M_{2KN}=\\&&\cr
&&\pmatrix{\phi_1&y_1&&&&&&&&&-z&0\cr 1&&&&&&&&&&0&z
\cr &&&y_{N-1}&-y_N&&&&&&&\cr &&1&\phi_N&0&y_N&&&&&&
\cr &&-1&0&-\phi_{N}&-y_{N-1}&&&&&& \cr&&&1&-1&&&&&&&
\cr&&&&&&&&&&&\cr &&&&&&&-y_1&-y_0&&&\cr
&&&&&&-1&-\phi_1&0&y_0&&\cr
&&&&&&-1&0&\phi_1&y_1&&\cr &&&&&&&1&1&&&
\cr &&&&&&&&&&&&\cr-y_0\over
z&0&&&&&&&&&&-y_1\cr 0&y_0\over
z&&&&&&&&&-1&-\phi_1}. \nonumber
\eea 

Following \cite{Boels:2003fh} let us define a matrix $G$ as
\be
G={\rm diag} (1,z^{1/2KN},z^{2/2KN},...,z^{(2KN-1)/2KN}) 
\ee
and map the matrix $M$ into ${\tilde M}$ using a similarity
transformation  $\tilde{M}=GMG^{-1}$. It can be seen that the obtained
matrix ${\tilde M}$ 
satisfies $S\tilde{M}S^{-1}=\tilde{M}$, where 
\be
S=\pmatrix{0&I_{2(K-1)N}\cr I_{2N} }
\ee 
is a matrix that generates a cyclic permutation of order $2N$ on $2KN$
elements. 
For the later use we note that the eigenvectors of $S$ are
$2KN$-dimensional space
which is a direct sum of $K$ $2N$-dimensional subspaces labeled by a different
$K-th$ root of unity $e^{2r\pi i\over 2N}$. Thus a basis for eigenvectors
of $S$
might be recast to $v_r^j=v^j(1,e^{2r\pi i\over 2N},e^{4r\pi i\over 2N},\cdots,
e^{(K-1)r\pi i\over 2N})$ for $j=1,2,\cdots, 2N$. 
 
Now the task is to show that the $M_{2KN}$ is somehow {\it related} to the
proper Lax matrix of the $SO(2KN-2K+2)$ gauge group. The invariance under
cyclic permutation ensures that the equations resulting from
$\tilde{M}_{2KM}$ collapse to those resulting from $M_{2N}$. In order to
find the spectral curve resulting from $M_{2KN}$, we note that
the matrix $\tilde{M}$ can explicitly be written as the following  
\be
\tilde{M}=\pmatrix{A&B&0&.&.&.&C\cr C&A&B&0&.&.&0\cr
0&C&A&B&.&.&.\cr.&.&.&.&.&.&B\cr B&0&0&0&.&C&A} 
\ee 
where
$A, B$ and $C$ are $2N\times2N$ matrices given by 
\bea
A=\pmatrix{\phi_1&{y_1\over z^{1/2KN}}&&&&&&0\cr
z^{1/2KN}&&&&&&&\cr &&&{y_{N-1}\over z^{1/2KN}}&-y_N\over
z^{1/KN}&&&& \cr&&&\phi_N&0&{y_N\over z^{1/KN}}&&\cr&&-z^{1/KN}
&0&-\phi_N&{-y_{N-1}\over z^{1/2KN}}&&
\cr&&&z^{1/KN}&&&&\cr&&&&&&&\cr&&&&&&&{-y_1\over
z^{1/2KN}}\cr 0&&&&&&&-\phi_1},\nonumber
\eea
\be
B=y_0z^{-1/KN}\pmatrix{0&.&.&0\cr.&.&.&.\cr-1&0&.&0\cr0&1&.&0},\;\;
C=z^{1/KN}\pmatrix{0&.&-1&0\cr0&.&0&1\cr.&.&.&.&\cr0&.&.&0}. 
\ee
Since the matrix $\tilde{M}$ commutes with $S$ one can use 
the basis of eigenvectors of $S$ to diagonalize it and thereby to
compute its determinant in terms of $A,B$ and $C$ matrices as the
following \cite{Boels:2003fh} 
\be 
\det(\tilde{M})=\prod_{t=1}^{K}\det(A+e^{2i\pi t\over
K}B+e^{-2i\pi t\over K}C)\;. 
\ee 
By making use of this relation, let us now evaluate the spectral curve 
coming from the {\it auxiliary Lax} matrix $\tilde{M}$. From the structure 
of matrix $\tilde{M}$ one finds
\be
\det(x{\bf 1}-\tilde{M})=\prod_{t=1}^K\left[P_{2N}(x)-x^2
\left(e^{{-2t\pi i\over K}}z^{{1\over t}}+\eta^2 e^{{2t\pi i\over K}}
z^{-{1\over t}}\right)\right]+x^{2K}
(z+{\lambda^{4KN-4K}\over z})=0\;,
\l{AXI}
\ee 
where $\eta$ is a $2K$-th root of unity. Note that
the $x$-dependent factor in front of the $z$ term comes from the fact that 
the matrix $\tilde{M}$ is made out of $K$ copies of the $SO(2N)$ Lax matrix 
in such a way that each one contributes a factor of $x^2$ and therefore 
altogether we get $x^{2K}$. We note also that the first term in the above 
expression, by
construction, is $z$-independent. So one can fix it with a proper
value for $z$. Indeed setting $z=-i|\Lambda^{2KN-2K}|$ the first term 
can be written as 
\bea
P_{2KN}(x)&=&\prod_{t=1}^K\left[P_{2N}(x)-x^2\eta\Lambda^{2N-2}
\left(e^{-i\pi{4t-1\over2K}}+e^{i\pi{4t-1\over2K}}\right)\right]\cr &&\cr 
&=&2^kx^{2K} \eta^K \Lambda^{2KN-2K}\prod_{t=1}^K\left[{P_{2N}(x)\over
2x^2\eta\Lambda^{2N-2}}-\cos\left(\pi{4t-1\over2K}\right)\right]\cr &&\cr
&=&
2x^K\eta^K\Lambda^{2NK-2K}T_K\left({P_{2N}(x)\over{2\eta
x^2\Lambda^{2N-2}}}\right)\;,
\eea 
where
$T_K(x)$ is the $K$-th Chebyshev polynomial of the first kind.
Plugging the final result into the auxiliary spectral curve ({\ref{AXI}) 
we get
\be
x^{K-2}\left\{2x^2\eta^K\Lambda^{2NK-2K}T_K\left({P_{2N}(x)\over{2\eta
x^2\Lambda^{2N-2}}}\right)+x^2(z+{\lambda^{4KN-4K}\over z})\right\}=0\;.
\ee
Since in general $x\neq 0$ we have
\be
P_{2KN-2K+2}(x)+x^2(z+{\lambda^{4KN-4K}\over z})=0
\ee
with 
\be
P_{2KN-2K+2}(x)=2x^2\eta^K\Lambda^{2KN-2K}T_K\left({P_{2N}(x)\over{2\eta
x^2\Lambda^{2N-2}}}\right) 
\ee 
which is exactly the spectral curve for $SO(2KN-2K+2)$ gauge group.

\section{SYM theory with $SO(2N+1)$ and $SP(2N)$ gauge groups }

\subsection{$SO(2N+1)$}
In this section we shall briefly study ${\cal N}=2$ SYM theory with gauge
group $SO(2N+1)$ on $R^3\times S^1$. We note, however, that since the 
group is nonsimply laced one will have to work with its dual
\cite{Martinec:1995by}
where the short loots and long roots are exchanged.
Therefore the integrable model one should deal with  
is a periodic Toda lattice based on the root system of dual gauge group
\cite{Martinec:1995by} which for the case of $B_N$ is given in terms of 
twisted Kac-Moody algebra $A^{(2)}_{2N-1}$.

The corresponding Lax matrix $M$ can be expressed as the following
\be
M=\sum_{i=1}^N(\phi_i\; h_i+y_i\;e_{\alpha_i}+e_{-\alpha_i})+z\;e_{\alpha_0}
+{y_0\over z}\;e_{-\alpha_0},
\l{Laxodd}
\ee
where $h_i, e_{\alpha_i}$ and $e_{\alpha_0}$ are generators of Cartan
subalgebra,
simple roots and affine root respectively. We note also that 
there is a constraint on
$y_i$'s given by $\prod_{j=0}^Ny_j\prod_{i=1}^{N-2}y_i=\Lambda^{4N-2}$.

Using this Lax matrix the Seiberg-Witten curve for the four dimensional
${\cal N}=2$
$SO(2N+1)$ SYM theory can be obtained from the spectral curve $\det(x{\bf
1}-M)=0$
which is\footnote{ The Seiberg-Witten curve for gauge group $SO(2N+1)$ was
first proposed in \cite{Danielsson:1995is}.} 
\be
x(z+{\Lambda^{4N-2}\over z})+{1\over 2}(x^{2N}+u_2x^{2N-2}+\cdots+u_{2N})=0\;.
\l{SOodd}
\ee

Let us now consider the theory on $R^3\times S^1$ with the following 
tree level superpotential 
\be
W={g_2\over 2}{\rm Tr}{\phi^2}+{g_4\over 4}{\rm Tr}{\phi^4}\;.
\l{ttt}
\ee
According to the proposal (\ref{pro}) the quantum superpotential is given by
\be
W={g_2\over 2}{\rm Tr}{M^2}+{g_4\over 4}{\rm
Tr}{M^4}+2L\log\left({\Lambda^{4N-2}\over
\prod_{j=0}^Ny_j\prod_{i=1}^{N-2}y_i}\right)\;,
\ee 
where $M$ is the Lax matrix (\ref{Laxodd}). Moreover we have also imposed
the constraint
on $y_i$ using a Lagrange multiplier $L$. This superpotential is a function of
$\phi_i, y_i$ and $L$. 

To be specific let us consider $SO(9)$ gauge group. By making use of the
explicit 
form of the Lax matrix $M$ and plugging it into the superpotential one
can read the equations of motion of $\phi_i$ as the following
\bea
W'(\phi_1)+g_4(2y_1\phi_1+y_0\phi_1+y_1\phi_2)&=&0,\cr
W'(\phi_2)+g_4(y_1\phi_1+2y_1\phi_2+2y_2\phi_2+y_2\phi_3)&=&0,\cr
W'(\phi_3)+g_4(y_2\phi_2+2y_2\phi_3+2y_3\phi_3+2y_4\phi_3+y_3\phi_4-
y_4\phi_4)&=&0,\cr
W'(\phi_4)+g_4(y_3\phi_3+2y_3\phi_4-y_4\phi_3+2y_4\phi_4)&=&0,
\eea
and for $y_i$'s one finds
\bea
g_2+g_4(2y_1+y_0+\phi_1^2)&=&2L\;y_0^{-1},\cr
g_2+g_4(y_2+y_1+y_0+\phi_1^2+\phi_2\phi_1+\phi_2^2)&=&2L\; y_1^{-1},\cr
g_2+g_4(y_1+y_2+y_3+y_4+\phi_2^2+\phi_2\phi_3+\phi_3^2)&=&2L\; y_2^{-1},\cr
g_2+g_4(3y_4+y_3+y_2+\phi_3^2+\phi_3\phi_4+\phi_4^2)&=&\;\;L\; y_3^{-1},\cr
g_2+g_4(y_4+3y_3+y_2+\phi_3^2-\phi_3\phi_4+\phi_4^2)&=&\;\;L\; y_4^{-1},
\eea
finally for $L$ we get $y_0y_1^2y_2^2y_3y_4=\Lambda^{14}$. These equations
can be solved to find the minimum of the corresponding supersymmetric
vacuum. For example
consider a solution such that $\phi_i=0$ for $i=1,2,3,4$. Then the equations 
for $y_i$'s can be solved leading to the following solution
\be
y_0=y_1=y_2=2y,\;\;\;\;\;y_3=y_4=y,\;\;\;\;\;y={1\over 2^{5/7}}\Lambda^2\;.
\ee
Plugging this solution into the superpotential one finds
\be
W=7\left(2^{2/7}g_2\;\Lambda^2+2^{4/7}{3\over 2}\;g_4\;\Lambda^4\right)\;,
\ee
which is exactly what we get from four dimensional field theory
\cite{Edelstein:2001mw} 
if we rescale $\Lambda$ as $\Lambda\rightarrow {1\over 2^{1/7}} \Lambda$.
This can,
for example, be seen from the Seiberg-Witten factorization (\ref{SWf}).
The same as that in the previous section this rescaling factor
can also be understood from the spectral curve (\ref{SOodd}). In fact we have 
again an extra factor of $1/2$ which can be absorbed in rescaling of $z$
and $\Lambda$
as $z\rightarrow z/2$ and $\Lambda\rightarrow \Lambda/2^{(2/4N-2)}$.

The supersymmetric vacua of $SO(2N+1)$ gauge theory with given superpotential
can also be lifted to the supersymmetric vacua of $SO(2NK-K+1)$ with the same
superpotential using the Chebyshev polynomials. This correspondence can also
be seen using its corresponding integrable model. In fact the situation is 
the same as that in the previous section, of course with the Lax matrix of
$SO(2N+1)$ gauge group. The final result is
\be
2x\eta^K\Lambda^{2NK-K}T_K\left({P_{2N}(x)\over{2\eta
x^2\Lambda^{2N-1}}}\right)+x(z+{\lambda^{4KN-2K}\over z})=0\;.
\ee

\subsection{$SP(2N)$}

Similarly one can also study the theory with gauge group $SP(2N)$. Being
a nonsimply
laced group points that we will have to work with its dual. In fact
in was shown \cite{Martinec:1995by} that the Seiberg-Witten curve for
this theory can be obtained from a periodic Toda lattice based on
the dual root system of $C_{N}^{(1)}$ which is given in terms of
twisted Kac-Moody algebra $D_{N+1}^{(2)}$. Its Lax matrix has the same form as
(\ref{Laxodd}) but, of course, written in terms of $D_{N+1}^{(2)}$ root system.
Moreover the constraint which should be imposed on $y_i$ now becomes 
$\prod_{i=0}^{N}y_i^2=\Lambda^{2N+2}$. Using this Lax matrix the spectral curve 
$\det(x{\bf 1}-M)=0$, reads
\be
(z-{\Lambda^{2N+2}\over z})^2+x^{2N}+u_2x^{2N-2}+\cdots +u_{2n}=0\;.
\ee

The same as in the previous section let us consider ${\cal N}=2\;SP(2N)$
SYM theory on $R^3\times S^1$ deformed by a tree level superpotential given
by (\ref{ttt}). Following the proposal (\ref{pro}) the quantum superpotential
reads
\be
W={g_2\over 2}{\rm Tr}{M^2}+{g_4\over 4}{\rm
Tr}{M^4}+2L\log\left({\Lambda^{2N+2}\over
\prod_{i=0}^Ny_i^2}\right)\;.
\ee 
Using the explicit form of Lax matrix we have evaluated the effective
superpotential
for the case of $SP(6)$. In the case of maximally confining vacuum we 
have found
\be
W=8(g_2\Lambda^2+{3\over 2}g_4\Lambda^4)\;.
\l{uuu}
\ee
To compare this with the four-dimensional field theory we could, for
example, study
the factorization of the corresponding Seiberg-Witten curve given by 
\cite{Argyres:1995fw}
\be
y^2=(x^2P_{2N}+2\Lambda^{2N+2})^2-4\Lambda^{4N+4}\;.
\ee
In general the factorization corresponding to the situation where gauge group 
is broken as $SP(2N)\rightarrow SP(2N_0)\times \prod_{i=1}^nU(N_i)$ is given by
\cite{Edelstein:2001mw}
\be
(x^2P_{2N}+2\Lambda^{2N+2})^2-4\Lambda^{4N+4}=x^2H_{2N-2n}F_{2(2n+1)}\;.
\ee
In the case of the maximally confining vacuum where $n=0$ the solution for this
problem is given in terms of Chebyshev polynomials
(see also \cite{Fuji:2002vv})
\be
P_{2N}(x)={2\Lambda^{2N+2}\over x^2}T_{N+1}\left({x^2\over 2\Lambda^2}-1\right)
-{2\Lambda^{2N+2}\over x^2}\;.
\ee
Plugging this solution to the above hyperelliptic curve one finds
\bea
y^2&=&\left[2\Lambda^{2N+2}T_{N+1}\left({x^2\over 2\Lambda^2}-1\right)\right]^2
-4\Lambda^{4N+4}\cr
&=&4\Lambda^{4N+4}\left[T_{N+1}^2\left({x^2\over
2\Lambda^2}-1\right)^2-1\right]\cr
&=&x^2\Lambda^{4N}(x^2-4\Lambda^2)U^2_{N}
\left({x^2\over 2\Lambda^2}-1\right)\;.
\eea
Form this solution one can read off the factorization points which can be 
found as the following 
\be
T_{N+1}\left({x_r^2\over 2\Lambda^2}-1\right)=0,\;\;\rightarrow\;\;
x^2_r=(2\Lambda)^2\cos^2\left({(r-{1\over 2})\pi\over 2(N+1)}\right)\;,
\ee
and therefore the gauge invariant parameters of the theory are given by
\be
\langle u_{2p} \rangle={N+1\over p}\pmatrix{2p\cr p}\Lambda^{2p}\;.
\ee
By making use of this solution the exact superpotential reads
\be
W=g_2 \langle u_{2} \rangle+g_4\langle u_{4}
\rangle=8(g_2\Lambda^2+{3\over 2}g_4
\Lambda^4)\;,
\ee 
in agreement with (\ref{uuu}).

Finally we note that the lifting from $SP(2N)$ to $SP(2N+2K-2)$ gauge theory
can also be done using the corresponding Lax matrices.

\section{Conclusions}

In this paper we have studied deformed ${\cal N}=2$ supersymmetric gauge 
theory with classical gauge group on $R^3\times S^1$. The deformation is 
given by a tree level superpotential. We have seen that the effective
quantum superpotential can be obtained by making use of an integrable
model which is related to the four dimensional theory. This integrable
model is the periodic Toda lattice which is based on the root lattice
of the affine Lie algebra corresponding to the gauge group. We note,
however, that for nonsimply laced gauge group we need to use its
dual which turns out to be twisted Kac-Moody algebra \cite{Adler2}.
This is, of course, what we expected from the relation between 
periodic Toda lattice and Seiberg-Witten models \cite{Martinec:1995by}.

The prescription to find the effective superpotential is simple and it 
just needs a substitution. More precisely if the tree level superpotential
is ${\rm Tr}W(\phi)$ then the effective superpotential is given by
${\rm Tr}W(M)$ where $M$ is the Lax matrix of the corresponding integrable 
model. As we see we do not even need performing any integrations. This might
be understood from the point that a $(4+d)$-dimensional gauge theory
could be studied by a $d$-dimensional auxiliary theory 
\cite{Dijkgraaf:2003xk}. Therefore in a three dimensional theory it seems
that no integrations are indeed needed. 

We note also that since the structure of the supersymmetric vacua 
on $R^3\times S^1$ is $R$ independent one could compare our results 
with that on $R^4$ theory. In fact we have shown that it precisely gives
the same vacuum structure of $R^4$ theory.
 
In our expressions for the effective superpotential a Lagrange multiplier 
is involved which has been used to impose the constraint on the parameters
of the integrable model. Following \cite{Boels:2003fh} this Lagrange 
multiplier can be identified with the glueball field $S$. Although in the
case where classically the gauge group remains unbroken one can integrate
in the glueball field $S$, for the situation where the gauge group is
classically broken this procedure remains unsolved. Moreover even 
in the first case it would be quite interesting to see how one can go
from $R^3\times S^1$ to $R^4$ getting for example the Veneziano-Yankielowicz
superpotential.

\end{document}